\documentclass{mem}
\usepackage{natbib}\usepackage{txfonts}\usepackage{balance}
\usepackage{graphicx}
\usepackage[a4paper]{hyperref}
\begin{document}
\def\teff{$T\rm_{eff }$}
\def\kms{$\mathrm {km s}^{-1}$}

\title{
MERLIN monitoring of recent core-collapse supernovae
}
\subtitle{}

\author{M.K. Argo
, R.J. Beswick, T.W.B. Muxlow, A. Pedlar, D. 
Fenech and H. Thrall}

\offprints{M.K. Argo}

\institute{
Jodrell Bank Observatory, University of Manchester, Macclesfield, Cheshire, 
SK11 9DL, UK.
\email{mkargo@jb.man.ac.uk}
}

\authorrunning{Argo }

\titlerunning{MERLIN monitoring of SNe}


\abstract{
The star formation rate (SFR) in starburst galaxies can be measured by many 
methods, one of which is through the supernova rate.  Due to the heavy 
dust obscuration in these galaxies, searches for new supernovae in the 
optical or infra-red can easily miss events occurring in the central 
starburst regions.  As part of a long term program to estimate the SFR in a 
sample of nearby starbursts we are using MERLIN and the VLA to regularly 
observe the galaxies for new radio supernovae.  As part of this project, 
regular MERLIN observations have been made of two recent optically bright 
supernovae: 2004dj and 2004et.  Both supernovae are of Type II and have been 
monitored frequently over periods of a few months, resulting in well sampled 
radio "light" curves for both objects.
\keywords{supernovae:individual:2004dj, 2004et -- galaxies:ISM -- 
galaxies:starburst}
}
\maketitle{}


\section{The monitoring program}	

The supernova (SNe) rate is a useful indicator of the star formation rate 
(SFR) in starburst galaxies.  In active star forming regions with large gas 
reservoirs it is likely that large numbers of massive stars will form.  If 
all stars with a mass greater than 8\,M$_{\odot}$ become supernovae via 
core collapse then, assuming a reasonable initial mass function, the SFR 
can be calculated as a simple function of the supernova rate.  Searches 
for new SNe are often carried out optically or in the infra-red which, 
although able to pick out explosions in the outer parts of starburst 
galaxies, can miss events in the heavily dust obscured central regions.

In order to observe new supernova events in the optically obscured centres 
of these galaxies, and hence estimate the SFR, we are using a combination 
of the NRAO's Very Large Array (VLA, \citealt{thompson80}) and the 
Multi-Element Radio Linked Interferometer Network (MERLIN, 
\citealt{thomasson86}) to regularly monitor a sample of ten starburst 
galaxies over five years.  Regular observations will allow the detection and 
follow-up of new radio supernovae (RSN) and supernova remnants (SNR), while 
the long timescale of the project will allow the flux evolution of existing 
remnants to be measured.  This will allow estimates of the star formation 
rates to be determined, leading to a comparison with the rates determined 
through other indicators.

The observing strategy involves observations of the ten galaxies in the 
sample, all of which are within 15 Mpc, roughly three times a year.  The VLA 
is used when in A or B configuration as reasonable resolution is required 
in order to separate new events from the rest of the galactic emission, 
and MERLIN is used when the VLA is in the more compact C or D configurations.  
Observations need not be more frequent than this as radio emission tends to 
rise later than at optical wavelengths and persists for several months in 
the case of core collapse supernovae (Type Ib/c and Type II SNe 
\citealt{weiler02}).  Note that Type Ia SNe are not radio bright to the 
detection limit of the VLA.

The first observation in this program was carried out using the VLA in B 
configuration in November 2003.  Several epochs have now been completed, with 
both the VLA and MERLIN, and observations of several supernovae have been 
made.


\section{Results}


\subsection{J103851+532927}

The first result from this program was a previously undiscovered RSN situated 
one arcminute from the centre of the nearby starburst galaxy NGC\,3310 
\citep{argo04a}.  This object (known as J103851+532927) is coincident with a 
group of H{\sc ii} regions and was visible in archive data as far back as 1986 
although, in the early 1990s and before, the flux was at least a factor 
of five lower.  The object is also coincident with a weak X-ray source, 
although no optical counterpart has been discovered despite searches through 
archival data.

The source had been noted in previous radio maps of the galaxy (e.g. 
\citealt{kregel01}), although it was assumed to be a background quasar.  The 
source displays properties which are uncharacteristic of normal quasar 
behaviour, however.  Firstly the source has a steep spectral index 
($\alpha \sim -$1.6 where S$\propto\!\!\nu^{+\alpha}$) and secondly it 
displays large flux variability with a sharp rise, followed by a slower 
decrease.  In the mid-1990s the flux at 1.4\,GHz increased by at least a 
factor of five to almost 10\,mJy before decreasing at approximately 10 per 
cent per year.

Since the discovery of this object, several more observations have been 
performed using MERLIN, and the resulting radio light curve of this object was 
published in \cite{argo04a}.  The fact that the SNe appears to have occured in 
the late 1990s so the radio emission has now persisted for over 5000 days 
implies that this was a Type II SNe.  If the source is at the distance of 
NGC\,3310 then the approximate peak 5\,GHz luminosity was $\sim 3 \times 
10^{19}$ W\,Hz$^{-1}$ which, although at the low end of the scale, is 
consistent with luminosities measured for other Type II SNe (see e.g. 
\citealt{weiler02}).


\subsection{2004dj}

\begin{figure*}[t!]
\resizebox{\hsize}{!}{\includegraphics[clip=true]{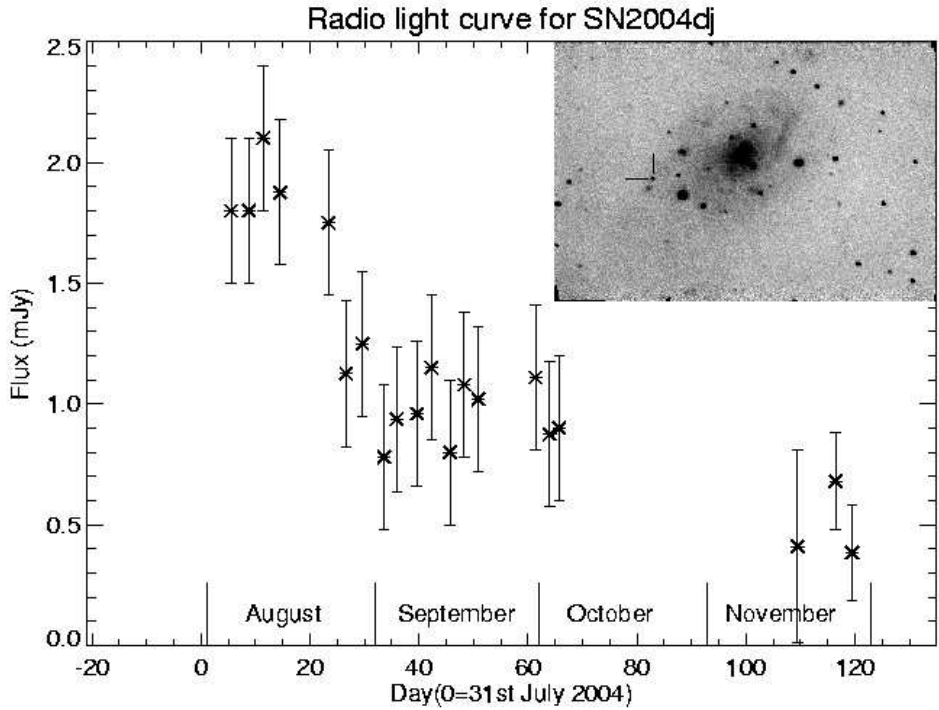}}
\caption{\footnotesize
Radio light curve for SN 2004dj from MERLIN 4994\,MHz observations.  
Observations from day 0 to early October were made with a single 217\,km 
baseline (Cambridge-Defford).  Each flux measurement is a 2.5 day 
vector-averaged point.  Subsequent observations were performed using the full 
MERLIN array.  The inset shows an optical image (4$\times$120\,s exposures) 
of the host galaxy NGC\,2403 obtained with a 10-inch Schmidt-Cassegrain 
telescope on April 2nd 2005, SN 2004dj is marked.}
\label{2004dj}
\end{figure*}

At the end of July 2004 a bright supernova was discovered optically in 
the nearby starburst galaxy NGC\,2403 \citep{nakano04}.  Peaking at a 
magnitude of 11.2, this was the brightest supernova seen in a decade.  MERLIN 
observations using a subset of the array began in early August 2004 and 
continued to early October, followed by imaging runs using the full array in 
November and December 2004.  This allowed a detailed 5\,GHz light curve to be 
determined, see Fig. \ref{2004dj} \citep{beswick05}.

The MERLIN observations allowed the position of the source to be determined 
to an accuracy of better than 50\,mas \citep{argo04b}, coincident with the 
optical \citep{nakano04} and $Chandra$ X-ray \citep{pooley04} positions, and 
the star cluster n2403-2866 \citep{larsen99}.  This illustrates the usefulness 
of MERLIN for this kind of program as, at the time, the VLA was in the most 
compact D configuration.  Although 2004dj was observed with the VLA 
\citep{stockdale04}, the measured position was affected by extended emission 
from the galaxy and lies $\sim$1\farcs2 from other measurements.

Weiler et al. (2002) find a relationship between the peak luminosity of the 
prompt emission from a Type II supernova at 6\,cm, L$_{\rm 6cm\,peak}$,
and the delay between explosion and peak luminosity of the form
\begin{equation}
{\rm L}_{\rm 6cm\,peak} \simeq 5.5^{8.7}_{3.4} \times 10^{16} (t_{\rm
6cm\,peak} - t_{o})^{1.4 \pm 0.2} {\rm W\,Hz}^{-1}
\end{equation}
where $t_{\rm 6cm\,peak}-t_{\rm o}$ is the delay measured in days.  For 
2004dj, assuming a peak flux density of 1.9$\pm$0.1\,mJy (equivalent to 
L$_{\rm 6cm peak} \approx  2.45 \times 10^{18}$ W\,Hz$^{-1}$ at the 
distance of NGC\,2403, making this one of the least luminous radio supernovae 
ever detected) implies $t_{\rm 6cm peak} - t_0 = 15^{+42}_{-10}$ days.  This 
places the date of explosion between 11 July and the date of optical 
detection, 31 July, broadly coincident with spectroscopic observations 
reported by \cite{patat04} which put the age of the supernova at 
approximately three weeks on August 3rd.

This SNe was optically classified as Type II-P \citep{patat04}, a relatively 
common type of supernova optically, but rarely detected at radio wavelengths.  
In fact, prior to 2004dj, the only two Type II-P SNe detected by radio 
telescopes were SN 1999em \citep{turtle87} for which no light curve was 
established, and the well-observed SN 1987A \citep{pooley02}, both of which 
were also relatively weak radio emitters.


\subsection{2004et}

In September 2004 another bright supernova was discovered, SN 2004et 
\citep{zwitter04}.  The host galaxy, NGC\,6946, is an active starburst in 
the monitoring sample, and also contains many historical supernovae.  As 
part of the monitoring program there were three recent observations of 
this galaxy, none of which showed any emission at the position of SN 
2004et when re-examined.  MERLIN observations began in early October and 
continued in parallel with observations of 2004dj until December 
\citep{beswick04b}.  The resulting radio light curve is shown in Figure 2.

\begin{figure*}[t!]
\resizebox{\hsize}{!}{\includegraphics[clip=true]{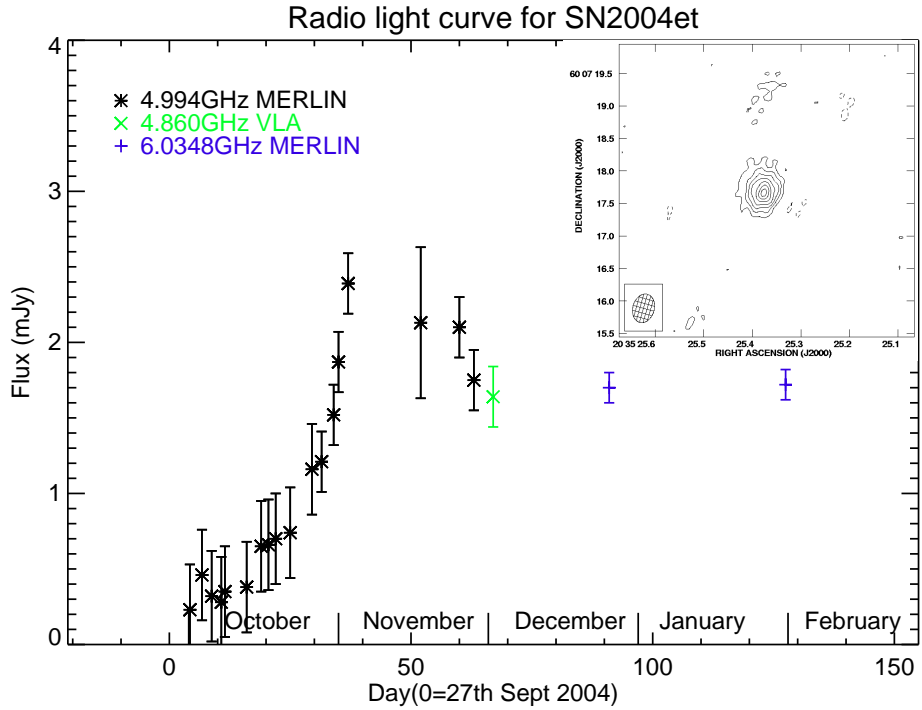}}
\caption{\footnotesize
Radio "light" curve for SN 2004et from MERLIN and VLA observations.  These 
MERLIN observations were also carried out using a subset of the array during 
October and early November.  Points from mid-November onwards are from imaging 
runs with the full MERLIN array.  The mid-November point was affected by bad 
weather.  The final November point is from a scheduled monitoring observation 
with the VLA.  The December and January points were made after the MERLIN 
frequency change to 6.035\,GHz.  The contour plot (inset) shows the MERLIN 
map from an observation on February 9th at 1.6\,GHz (see text).  Contours are 
$-$1, and 1 to 10 $\times$ 0.2\,mJy/beam.
}
\label{2004et}
\end{figure*}

Observations at 5\,GHz began while the radio emission was still 
increasing, with the peak ($\sim$2.5\,mJy) occuring sometime in early 
November.  Assuming a distance to NGC\,6946 of 5.5\, Mpc \citep{pierce94} 
gives a peak 5\,GHz luminosity of $\sim8.7\times10^{18}$\,W\,Hz$^{-1}$, 
3.5 times more luminous than 2004dj.

For 2004et, equation (1) results in $t_{\rm 6cm\,peak} - t_{\rm o}$ of
$37^{+108}_{-24}$ days.  The peak 6\,cm flux was actually measured on day 37 
after initial detection, although in this case there is no spectroscopy 
which gives an estimate of the explosion date.  Unfortunately there is a 
gap in the MERLIN observations of two weeks after the measurement on day 37.  
The flux could have continued to increase during this period, so there is some 
additional uncertainty in this figure.

This source was also observed with MERLIN in February 2005 at a frequency 
of 1.658\,GHz and was detected with a flux of 1.8$\pm$0.2\,mJy.  VLA monitoring 
observations in April 2005 detected the source with a flux of 1.5$\pm$0.1\,mJy at 
1.425\,GHz and only marginally at 4.860\,GHz, although this point is 
heavily affected by bandwidth smearing.


\subsection{Other supernovae}

{\bf 2004am:} Another event which has been observed with MERLIN as part of 
this program was the supernova candidate 2004am in M82, reported as an 
infra-red detection early in 2004 by \cite{mattila04a}.  Observations of M82 
have been carried out regularly as part of the monitoring program but, as yet, 
no radio emission has yet been detected from this object to a 3-$\sigma$ 
limit of 0.18 mJy\,bm$^{-1}$ at 5\,GHz \citep{beswick04a}.

{\bf 2004gt: } This Type {\rm I}b supernova was reported in NGC\,4038 in 
December 2004 \citep{monard04}.  This galaxy is at too low a 
declination for reliable observations with MERLIN although, as the system is 
in the monitoring sample, VLA observations were obtained on November 1st (42 
days before the date on which it was discovered optically) and March 17th (95 
days after the optical detection).  The 5\,GHz VLA maps from both epochs show 
some emission at the position of the source, but nothing above the existing 
H{\sc ii} region \citep{neff00}.  Maps were also obtained at 8.4 and 15\,GHz 
but no emission was detected at the coordinates of 2004gt.

{\bf 2005V: } Reported in January \citep{mattila04b}, this event occured in 
another galaxy in our sample, NGC\,2146.  MERLIN observations were made in 
April at 1.6\,GHz and in early February at 6\,GHz, but no significant 
emission (3-$\sigma$ limits: 142 and 207 $\mu$Jy respectively) has yet been 
observed.


\section{Future}

In the near future we intend to continue monitoring of both 2004dj and 
2004et.  Although the fluxes of both RSN have now decreased significantly, it 
is important to keep observing both objects in the long term in order to 
detect the onset of emission from the subsequent remnants.
In the longer term, regular observations of the galaxies in the sample will 
continue, as well as observations of other optically discovered supernovae 
with MERLIN when practical.

As a by-product of the monitoring program, the large volume of data collected 
will also allow a thorough investigation of long-term variability of the 
compact components in each galaxy.

The observations so far show the advantages of using MERLIN compared to the 
VLA which is in compact configurations for approximately half the year.  
Programs like this will become much less time consuming when {\em e}-MERLIN 
\citep{garrington04} comes on line, as observations which currently take 
several hours will require much less time due to the dramatic increase (a 
factor of $\sim$30) in sensitivity.


\begin{acknowledgements}
MERLIN is run by the University of Manchester as a National Facility on
behalf of PPARC.
The National Radio Astronomy Observatory is a facility of the National
Science Foundation operated under cooperative agreement by Associated
Universities, Inc.
MKA acknowledges support from a PPARC studentship.
Thanks to A. Rushton and N. Vaytet for the optical image of NGC\,2403 taken at JMOO, 
Manchester.
\end{acknowledgements}


\bibliographystyle{aa}


\end{document}